# Frequency Dependence of Rotor's Free Falling Acceleration and Inequality of Inertial and Gravity Masses

Alexander L. Dmitriev

*St-Petersburg State University of Information Technologies, Mechanics and Optics,
49 Kronverksky prospect, 197101, St-Petersburg, Russia, (812)3154071
alex@dmitriyev.ru*

**Abstract.** Results of measurements of free falling acceleration of a closed container with a rotor of a mechanical gyroscope placed inside it on the frequency of the rotor rotation are briefly described. Time of separate accelerations measurements is 40 ms, the period of sampling is from 0.5 up to 1.0 minute. In rotation's frequencies range of 20-400 Hz, the negative changes of free falling container acceleration prevail. On individual frequencies the "resonant" maxima and minima of acceleration are observed. The obtained data apparently contradict the equivalence principle of inertial and gravitating masses. The expediency of development of ballistic gravimetry of high time resolution with use of rotating or oscillating test bodies is noted.

From Galilee's times the measurements of free falling acceleration (FFA) of bodies are one of the main techniques of quantitative determination of gravitation properties. The accuracy of modern gravimeters comes up to nm/s$^2$ units and long ago the gravimetry has become a basis of experimental geophysics and a geodesy [1]. High sensitivity of ballistic gravimeters is basically provided due to two circumstances. First, thermal and mechanical stability of gravimeters design and their components, including test bodies. Second, longer time of integration and bigger number of registered gravimeteric data (in the best laser gravimeters the observations are conducted daily and the number of samples runs into thousands). In these measurements it is certainly supposed that parameters of the gravitational field of the Earth are constant, at least during the time of measurements, and that quite often the observed big scattering of measured samples values of FFA is caused by measurement errors, geophysical and technical factors, or artifacts. The results obtained over long observation time periods, though they give the record values of accuracy in measurements of FFA average values, are uninformative for researches of short-time changes of the gravitation field. Such changes can be caused by both external astronomical influences and complex physical processes in the core and volume of the Earth.

The high time resolution gravimetry , for example, at the level of the 10th - 100th fractions of a second, will give some valuable information on non-stationary geophysical processes, including interrelation of gravitational and magnetic fields of the Earth.

It is necessary to note that in ballistic measurements of FFA the physical state of a test body is of the basic importance. Movement of an accelerated, caused by external elastic (electromagnetic in nature) forces a body or its microparticles affects the measured mass of the body and the acceleration of its free fall. Such an "active" state of a test body is created during its heating (increase of intensity of chaotic movement of its microparticles), rotation, oscillations and impact effects [2-5]. Measurements of acceleration of rotor free falling of a mechanical gyroscope were usually made with the purpose of verification of «equivalence principle» (the review of the appropriate publications is given in [6]). As a rule, in these experiments the axis of a rotor was positioned vertically, the high accuracy of measurements was achieved by the big number of sample data, and measurements of FFA in a narrow range of frequencies of rotor rotation were carried out.

Measurements of free falling acceleration (or mass) of rotor with the horizontal axis of rotation are interesting due to the fact that in doing so the accelerated movement of material particles of the rotor is directed not across but along the vector of gravity force. Just in these conditions, the display of "nonclassical" properties of gravitation can probably appear, including effects which can be considered as analogues of phenomena of Faraday's law induction and Lentz's rules known in electrodynamics [7-9].

In our experiment we measured the free falling acceleration of the magnetically-, thermally- and sound-isolated container with a vacuumed aviation rotor inside it [10]. The maximal rotation frequency of a rotor is 400 Hz, the run out time of rotor is 22 min. Fall path length of the container is 30 mm, readout time of sample value of gravity acceleration is near 40 ms, the period of sampling is from 0.5 up to 1.0 minutes. The principle of measurements is based on photoregistration of movement of the scale in form of three horizontal strings fixed on the container. At the maximal falling velocity of the container equal to 60 cm/s and its dimensions of 82x82x66 mm, the joint influence of buoyancy and resistance force of air in FFA measurements did not exceed 0.1 cm/s$^2$. The error of some measurements of the container FFA was within the limits of 0.3-0.6 cm/s$^2$ and was basically determined by accuracy of readout times of registration of pulse signals in movement of the scale (near 1 microsecond).

The example of experimental frequency dependence of FFA changes $\Delta g(f)$ (in Gal) of the container, containing a rotor with a horizontal rotation axis, is shown in the Figure.

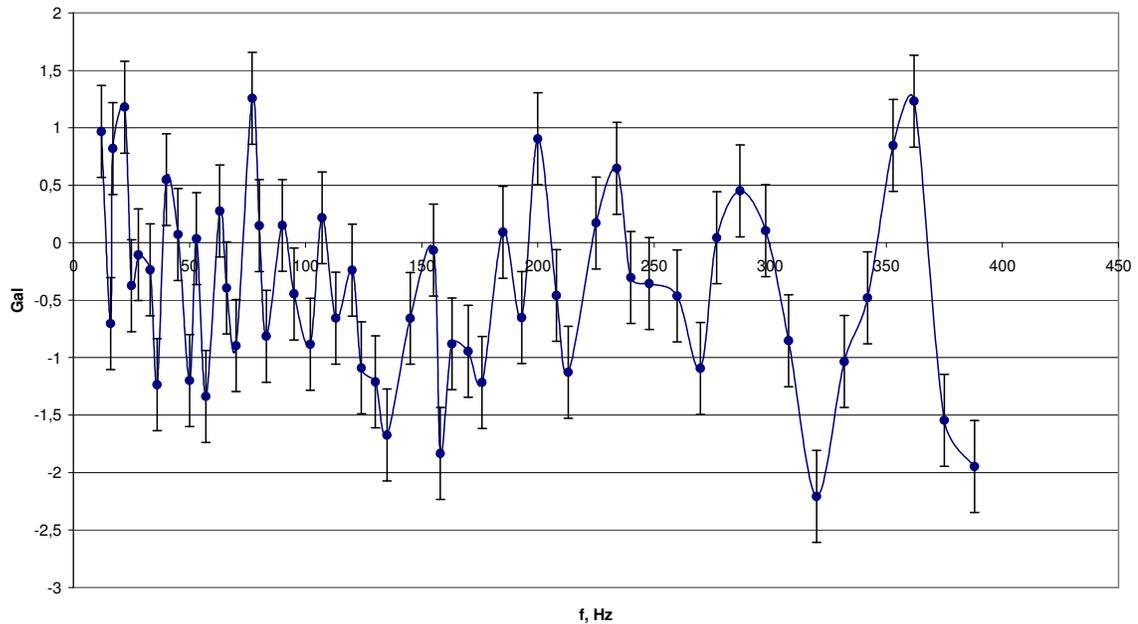

The value $\Delta g(0) = 0$ corresponds to acceleration of free falling of the container with a motionless rotor; FFA measurements of the container with a motionless rotor were carried out till the moment when rotor got going and after its run out time, in so doing the FFA values of the container, averaged by results of 10 measurements with a motionless rotor, coincided to the accuracy of 0.05%.

The features of frequency dependence of FFA changes are its stochastic character, narrow extremes which are the most appreciable near frequencies of 320 and 360 Hz, and prevalence, on the average, of the negative values of $\Delta g(f)$. These results do not contradict the earlier executed measurements of the FFA of the container with two rotors in which at frequencies of rotation 380-350 Hz the appreciable increase of FFA was noted [11]. The reduction tendency of size of the change $\Delta g(f)$, averaged for run out time of the rotor, is also in agreement with the data of measurements of rotor weight carried out with use of a precision comparator [3]. Statistically significant prevalence of negative average values of $\Delta g(f)$ in tens of series of measurements was observed. The resonant character of negative changes of $\Delta g(f)$, clearly expressed in rotation frequencies near 320 Hz, also proved to be true in repeated experiments. Similar laws at vertical orientation of an axis of a rotor are observed which can apparently be explained by a nonzero vertical component of oscillations of particles of the massive body of a gyroscope in rotation of the rotor.

The "nonclassical" properties of gravitation are just found in dynamic experiments in which an influence on a test body of external, not gravitational effects is significant, and the deep interrelation of gravitational and electromagnetic fields is most clearly expressed. The development of high time resolution ballistic gravimetry with use of rotating or oscillating test bodies will contribute to a deeper understanding of physics of gravitation.

## Conclusions

1. The free falling acceleration of a container with a mechanical gyroscope rotor inside it, measured in period of time less than 0.05 s, considerably differs from normal acceleration of gravity and in the range of frequencies of rotor rotation equal to 20-380 Hz the difference of such accelerations achieves several units of cm/s$^2$.
2. The frequency dependence of change of free falling acceleration of the container (rotor) has stochastic, and at some frequencies of rotor rotation, for example, near 320 Hz, a resonant character. Both at vertical and horizontal orientations of rotor rotation axis, a reduction of free falling acceleration of the container with a rotating rotor prevails.
3. A change, including reduction (levitation) of free falling acceleration of a closed container with a rotating rotor, seems to contradict the principle of identity of inertial and gravitation masses of a body.
4. The ballistic gravimetric researches made with the high time resolution and using rotating or oscillating test bodies are informative in measurements of dynamic characteristics of the gravitational field of the Earth and also contribute to development of physics of gravitational interaction.